\documentclass[%
reprint,
 amsmath,amssymb,
 aps,
 prx,
]{revtex4-2}
\usepackage{booktabs}
\usepackage{amsmath}
\usepackage{amsfonts}
\usepackage{amssymb}
\usepackage{graphicx}
\usepackage{dcolumn}
\usepackage{multirow}
\usepackage{color}
\usepackage{bm}
\usepackage[hyperfootnotes=true,bookmarks=true,]{hyperref}

\hypersetup{ final,		
			colorlinks=true,
			linkcolor=red,
            linktoc=page,
			anchorcolor=black,
			citecolor=red,
			urlcolor=blue}

\usepackage{soul}
\usepackage{comment}

\usepackage[tickmarkheight=2pt,]{todonotes}

\usepackage[
margin=1.2in,
]{geometry}

\begin{document}

\preprint{PREPRINT NUMBER}

\title{Inference through innovation processes tested in the authorship attribution task}

\author{Giulio Tani Raffaelli}
 \altaffiliation[At the time of this study at ]{Sapienza University of Rome, Physics Department, Piazzale Aldo Moro 5, 00185 Rome, Italy}
\affiliation{%
Institute of Computer Science of the Czech Academy of Sciences, Pod Vodárenskou věží 2, 182 07 Prague, Czech Republic
}%
\author{Margherita Lalli}
 \altaffiliation[At the time of this study at ]{Sapienza University of Rome, Physics Department, Piazzale Aldo Moro 5, 00185 Rome, Italy}
\affiliation{
Scuola Normale Superiore, Piazza dei Cavalieri 7, Pisa, Italy}%
\author{Francesca Tria}%
 \altaffiliation[Also at ]{Complexity Science Hub (CSH) Vienna, A-1080 Vienna, Austria}
 \email{francesca.tria@uniroma1.it}
\affiliation{%
 Sapienza University of Rome, Physics Department, Piazzale Aldo Moro 5, 00185 Rome, Italy}%

\date{\today}

\begin{abstract}
\section*{Abstract}
Urn models for innovation 
capture fundamental empirical laws shared by several real-world processes. The so-called  urn model with triggering includes, as particular cases, the urn representation of the two-parameter Poisson-Dirichlet process and the Dirichlet process, seminal in Bayesian non-parametric inference. In this work, we leverage this connection to introduce a general approach for quantifying closeness between symbolic sequences and test it within the framework of the authorship attribution problem. The method demonstrates high accuracy when compared to other related methods in different scenarios, featuring a substantial gain in computational efficiency and theoretical transparency. 
Beyond the practical convenience, this work demonstrates how the recently established connection between urn models and non-parametric Bayesian inference can pave the way for designing more efficient inference methods. In particular, the hybrid approach that we propose allows us to relax the exchangeability hypothesis, which can be particularly relevant for systems exhibiting complex correlation patterns and non-stationary dynamics.

  \end{abstract}

\maketitle

\section{Introduction}
Innovation enters a wide variety of human activities and natural processes, from artistic and technological production to the emergence of new behaviours or genomic variants.
At the same time, the encounter with novelty  permeates our daily lives more extensively than we typically realise.
We continuously meet new people, learn and incorporate new words into our lexicon, listen to new songs, and embrace new technologies.
Although innovation and novelties (i.e., new elements at the individual or local level) operate at different scales, we can describe their emergence within the same framework, at least in certain respects~\cite{tria2014dynamics}.
Shared statistical features, including the well-known Heaps'~\cite{heaps1978information}, Taylor's~\cite{taylor_1961,gerlach2014,tria2018zipf,tria2020} and Zipf's~\cite{zipf35,moreno2016large} laws, suggest a common underlying principle governing their emergence. In this respect, an intriguing concept is  {the expansion into the adjacent possible}~\cite{kauffman_2000}. The adjacent possible refers to the set of all the potential innovations or novelties attainable at any given time. When one of these possibilities is realised,  the space of the actual enlarges, making additional possibilities achievable and thus expanding the adjacent possible.
The processes introduced in~\cite{tria2014dynamics} provide a mathematical formalisation of these concepts, 
extending Polya's urn model~\cite{polya_1930} to accommodate infinitely many colours.
They generate sequences of items exhibiting Heaps', Zipf's, and Taylor's laws. 
The most general formulation of the modelling scheme proposed in~\cite{tria2014dynamics}, the urn model with semantic triggering, also captures correlations in the occurrences of novelties, as observed in real-world systems.
Further generalisations have been explored to capture the empirical phenomenology in diverse contexts: network growth and evolution~\cite{Ubaldi_2021}, the varied destinies of different innovations~\cite{monechi2017waves}, and mutually influencing events~\cite{aletti2023interacting}. Additionally, the proposed modelling scheme can be cast within the framework of random walks on graphs, offering further intriguing perspectives and broadening its scope of applications~\cite{Latora2018,iacopini2020interacting,di2022social,di2023dynamics}.

We now want to address the question of whether these generative models can also be successfully used in inference problems. This question is further motivated by the precise connection that has been established~\cite{tria2018zipf,tria2020} between the urn models in~\cite{tria2014dynamics} and seminal processes in Bayesian nonparametrics.
The latter is a powerful tool for inference and prediction in innovation systems, where possible states or realisations are not predefined and fixed once and for all. 
Nonparametric Bayesian inference enables us to assign probabilities to unseen events and to deal with an ever-increasing number of new possibilities.
Various applications have been proposed in diverse fields, including (but not limited to) estimation of diversity~\cite{lijoi2007bayesian,favaro2009bayesian,chakraborty2019using,holec2019bayesian,masoero2022more}, classification problems~\cite{gershman2012tutorial,ni2020scalable}, Bayesian modelling of complex networks~\cite{schmidt2013nonparametric,hu2019variational}, 
and they take a considerable role in Natural Language Processing~\cite{teh2010hierarchical,blei2010nested}. 

The simplest model described in~\cite{tria2014dynamics}, the urn model with triggering (UMT), 
reproduces, with a specific parameter setting, the conditional probabilities that define the two-parameter Poisson-Dirichlet process~\cite{Pitman1997}, referred to as PD hereafter, that
 generalises the Dirichlet process~\cite{ferguson1973}. The PD and the Dirichlet processes have gained special relevance as priors in Bayesian nonparametrics due to their generality and manageability~\cite{de2013gibbs}, and the PD process
 predicts the Heaps, Zipf and Taylor laws, making its use more convenient in linguistically motivated problems. 
 
Here, we aim to explore the potential of the outlined connection between urn models for innovation and priors for Bayesian nonparametric inference. As a sample application,
we address the authorship attribution task~\cite{Fadel2020overview}.

The PD and Dirichlet processes have already been considered as  underlying models for natural language processing and for authorship attribution purposes.  The proposed procedures  interpret the outputs of PD (or Dirichlet) processes as sequences of identifiers for distributions over words (i.e., topics)~\cite{Blei2010ProbabilisticTM} and
measure similarity among texts or authors based on topics' similarity~\cite{seroussi2014authorship,yang2017authorship}. We briefly discuss topic models in the Methods section. It is  worth  stressing here that these approaches have led to hierarchical formulations that require efficient sampling algorithms for solving the problem of computing posterior probabilities~\cite{Blei2010ProbabilisticTM,blei2010nested,teh2006collapsed,porteous2008fast}.  
 Moreover, these methods strongly rely on exchangeability,
 mainly due to the property of conditional independence it implies, through the de Finetti and Kingman theorems~\cite{deFinetti1937,kingman1978a}, and for guaranteeing the feasibility of the Gibbs sampling procedure~\cite{teh2010hierarchical,blei2010nested}.
Exchangeability refers to the property of the joint probability of a sequence of random variables  being invariant under permutations of the elements. Notwithstanding the powerful tools it provides, this assumption is often unrealistic when modelling real-world processes.

We take a different perspective by interpreting the outputs of the underlying stochastic processes directly as sequences of words in texts or, more generally, tokens. Language serves as a paradigmatic example where novelty enters at different scales, ranging from true innovation—creation and diffusion of new words or meanings—to what we denote as novelties—the first time an individual adopts or encounters (or an author uses in their production) a word or expression.
We thus borrow from information theory~\cite{shannon1948mathematical, Cover_Thomas_2006} the conceptualization of a text as an instance of a stochastic process and consider urn models for innovation processes as underlying generative models. Specifically, we here consider the UMT model in its exchangeable version, which is equivalent to the PD process.  We opt out of a fully Bayesian approach and use a heuristic method to determine the base distribution of the process—that is, the prior distribution of the items expected to appear in the sequence.

The overall change in perspective we adopt allows us to avoid the Monte Carlo sampling required in hierarchical methods. Moreover, while we consider here an exchangeable model, exchangeability is not crucial in our approach, 
paving the way for an urn-based inferential method that considers time-dependent correlations among items.

When comparing our method to various approaches used in authorship attribution tasks, we find promising results across different datasets (ranging from literary texts to blogs and emails), demonstrating that the method can scale to large, imbalanced datasets and remains robust to language variation.

\section{Results}
\subsection{The authorship attribution task} \label{sec:aa}

To demonstrate a possible application of the UMT generative model for an inference problem,
we used the probabilities of token sequences derived from the process to infer the authorship of texts. 
In the authorship attribution task, one is presented with a set of texts with known attribution -- the reference corpus -- along with a text $T$ from an unknown author. The goal is to attribute $T$ to one of the authors represented in the corpus (closed attribution task) or more generally, to recognise the author as one of those represented in the corpus or possibly as a new, unidentified author (open attribution task)~\cite{stamatatos2009survey}. 
Here we explicitly consider the case of the closed attribution task, although several strategies can be adopted to apply the method  in open attribution problems as well.

Following the framework of Information Theory~\cite{shannon1948mathematical,Cover_Thomas_2006}, we can think of an author as a stochastic source generating sequences of characters. In particular, a written text is regarded as a sequence of symbols, which can be dictionary words or, more generally, short strings of characters (e.g., $n$-grams if such strings have a fixed length $n$), with each symbol appearing multiple times throughout the sequence. Each symbol constitutes a novelty the first time it is introduced.

We evaluate the similarity between two symbolic sequences by computing the probability that they are part of a single realisation from the same source.
More explicitly, let $x_1^n$ and $x_2^m$ be two symbolic sequences with length  $n$ and $m$ respectively.
Given their generative process---their source---we can compute the conditional probability $P(x_1^{n} \mid x_2^{m})$, that is, the probability that $x_1^n$ is the continuation of $x_2^m$. 
In the authorship attribution task, the anonymous text $T$ is represented by a symbolic sequence $x_T$, while an author $A$ by the symbolic sequence $x_A$ obtained by concatenating the texts of $A$ in the reference corpus.
It is worth noting that an author $A$  affects the probability of $T$ both by defining the source and through the sequence $x_A$. We will use the notation 
 $P(T\mid A) \equiv P_A(x_T \mid \! x_A )$ for the conditional probability of $T$ to continue the production of $A$.
The anonymous text $T$ is  attributed to the author $\tilde{A}$ that maximises such conditional probability: $ \tilde{A} = \max\limits_{A} P(T \mid \!A )$. 
We thus need to specify the processes generating the texts and the elements $x_i$  of the symbolic sequences, i.e., the tokens. 

\subsubsection{The tokens}
We can make several choices for defining the variables---or tokens---$x_i$. In what follows, we consider two alternatives:
first, we consider
 Overlapping Space-Free $N$-Gram~\cite{koppel2011authorship} (OSF). 
These are strings of characters of fixed length $N$ as tokens, including spaces only as the first or last characters,
thereby discarding words shorter than $N-2$.  This choice has often yielded the best results.
 Secondly, we explore a hybrid approach where we exploit the structures captured by the Lempel and Ziv compression algorithm (LZ77)~\cite{ziv1977universal}. We define LZ77 sequence tokens as the repeated sequences extracted through a modified version of the Lempel and Ziv algorithm,  which has been previously used for attribution purposes~\cite{lalli2018ferran}. 
For each dataset, we select the token specification that provides the best performance. In the Supplementary  Results, we compare the achieved accuracy when using the token definitions discussed above as well as when using simple dictionary words as tokens.

\subsubsection{The generative process and the posterior probabilities}\label{sec:process_and_prob}
We consider the UMT model in its exchangeable version, which provides an urn representation of the PD process. The latter is defined by the
conditional probabilities of drawing at time $t + 1$ an { old }(already seen) element $y$   and a {new} one (not seen until time $t$). They are given, respectively, by:
 \small
\begin{equation}\label{eq:pPD}
    \begin{split}
&P(x_{t+1}= y  \mid x^t) =  \frac{n_{y,t}\!-\alpha}{\theta\!+\!t} ,\,\,\, \text{if} \,\,\, n_{y,t}>0 \\
&P(x_{t+1}= y  \mid x^t) =  \frac{\theta\!+\!\alpha D_t}{\theta\!+\!t}P_0(y) , \,\,\, \text{if} \,\,\, n_{y,t}=0 
\end{split}
\end{equation}
\normalsize
where \(n_{y,t}\) is the number of elements of type  $y$ at time $t$
and 
$D_t$ is the total number of distinct types appearing in $x^t$;  $0< \alpha < 1$ and $\theta > -\alpha$ are two real-valued parameters and $P_0(\cdot)$ is a given distribution on the variables' space, called the  base distribution.
The UMT model does not explicitly define the prior probability for the items' identity, i.e., the base distribution $P_0$. 
The latter can be independently defined on top of the process, in the same way as for the Chinese restaurant representation of the Dirichlet or PD processes~\cite{pitman2006combinatorial} (please refer to section UMT and PD processes
in the Methods for a thorough discussion on the urn models for innovation and their relation with the PD process). 

Crucially,  Eqs.~\eqref{eq:pPD}  are only valid when $P_0$ is non-atomic, which implies that each new token can be drawn from $P_0$ at most once with probability one. 
 On the contrary, when $P_0$ is a discrete probability distribution (it has atoms), an already seen value $y$ can be drawn again from it, and the conditional probabilities no longer have the simple form shown in Eqs.~\eqref{eq:pPD} (as detailed 
in the Methods).
In a problem of language processing, the tokens are naturally embedded in a discrete space, which has led to the development of hierarchical formulations of the PD process~\cite{Teh2006a,Teh2006b}. In these approaches, the  $P_0$ is the (almost surely) discrete outcome of another PD process with a non-atomic base distribution. 
Here we follow a different approach. We regard $P_0$  as a prior probability on the space of new possibilities. In this view, the tokens take values from an uncountable set, and thus the probability of drawing the same token $y$ from $P_0$ more than once is null. 
As a consequence, we can use the simple Eqs.~\eqref{eq:pPD},  where we need to make some arbitrary choices for the actual definition of  the base distribution. 
In the following, we identify $P_0(y)$ with the frequency of $y$ in each dataset, while still treating $P_0$ as a non-atomic distribution by ensuring that each item can be drawn at most once from it. However, this raises a tricky question of normalisation, which strongly depends on the dataset, resulting in the arbitrary modulation of the relative importance of innovations and repetitions.
We have addressed this problem heuristically by introducing an additional parameter $\delta>0$ that multiplies $P_0$: it suppresses ($\delta<1$) or enhances ($\delta>1$) the probability of introducing a novelty in $T$.
In addition, we consider an author-dependent base distribution by discounting the vocabulary already appearing in $A$ (details are given in section The strategy of $P_0$
in the Methods section). 
To summarise,  the conditional probabilities $P(T \mid \!A )$ are derived from Eqs.~\eqref{eq:pPD}, where the base distribution $P_0(y)$ is defined as discussed above.  Different values of $\alpha$ and $\theta$ characterise the specific distribution associated with each author.  We fix \(\alpha_A\) and \(\theta_A\) for each author $A$ to the values that maximise her likelihood (refer to the Supplementary  Methods for details). 
We denote by $D_K$ (with $K=A,T$) the number of types (i.e., distinct tokens) in  $A$ and $T$, and by $D_{T\cup A}-D_A$ the number of types in $T$ that do not appear in $A$.
The conditional probability  of a text $T$ to be the continuation of the production of an author $A$ reads:
\begin{eqnarray}\label{eq:condp}
&& P(T \mid \!A ) =\frac{(\theta_{A} + \alpha_{A} D_{A}   \mid \alpha_{A})_{D_{T\cup A} -  D_{A} }}{(\theta_{A} + m)_{n}}\prod_{j=1}^{D_{T}} Q_j , \nonumber  \\
 && \nonumber \\
 && Q_j  \equiv \begin{cases}
      (1-\alpha_{A})_{n^T_{j}-1}P_0(y_j) & \text{if }y_j \notin A \\
      (n^A_{j}\!-\!\alpha_{A})_{n^T_{j}} & \text{otherwise.}
    \end{cases}
 \end{eqnarray}
where   $n^K_{j}$ is the number of occurrences of $y_j$ in $K$ (with $K=A,T$), such that $\sum_j n^A_{j} = m$ and $\sum_j n^T_{j} = n$.
The Pochhammer symbol and  the Pochhammer symbol with increment $k$ are defined respectively by $(z)_n \equiv z(z+1)\ldots (z+n-1)=\Gamma(z+n)/\Gamma(z)$  and $(z\mid k)_n \equiv z(z+k)\ldots \left(z+(n-1)k\right)$.

In practice, when attributing the unknown text, we adopt the procedure of dividing it into fragments and evaluating their conditional probability separately. The entire document is then attributed either to the author that maximises the probabilities of most fragments or  to the author that maximises the whole document probability computed as a joint distribution over independent fragments (i.e., as a product of the probabilities of its  fragments).
We optimise this choice for each specific dataset, as described in the Supplementary  Methods.

\begin{figure*}
\includegraphics[width=1\linewidth]{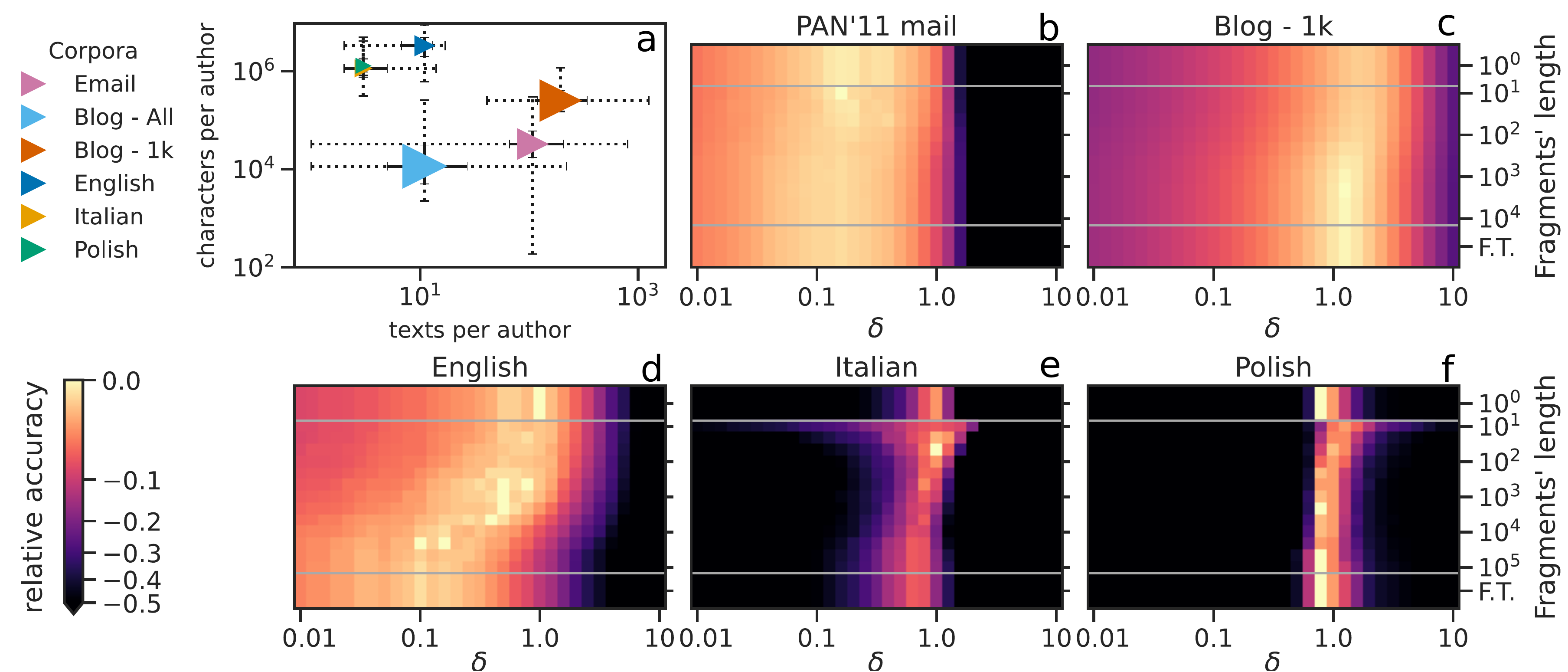}
\caption{\textbf{Corpora sizes and the impact of model parameters on attribution accuracy.} In panel (a) we offer a pictorial view of various characteristics related to the size of the considered corpora. 
The size of the triangles is proportional to the logarithm
 of the corpus size, measured as number of documents.  In the $x$ and $y$ axes we represent for each corpus  the distribution of the numbers of texts ($x$ axis) and of the numbers of  characters  ($y$ axis) per author. Specifically, 
 the continuous line bars represent the interquartile range of the distributions and the dotted lines show the 95\% interval, to highlight their long tails. Panels  (b) to  (f) report the attribution accuracy varying the length of the fragments and the $\delta$ value. The colour scale refers to the difference relative to the maximum attribution accuracy obtained in each dataset.
In the upper band, the considered length of fragments is of a single token. In the lower band, the text is not partitioned in fragments (Full Text).}
\label{fig:datasets}
\end{figure*}

 \begin{figure}[ht]
\includegraphics[width=1\linewidth]{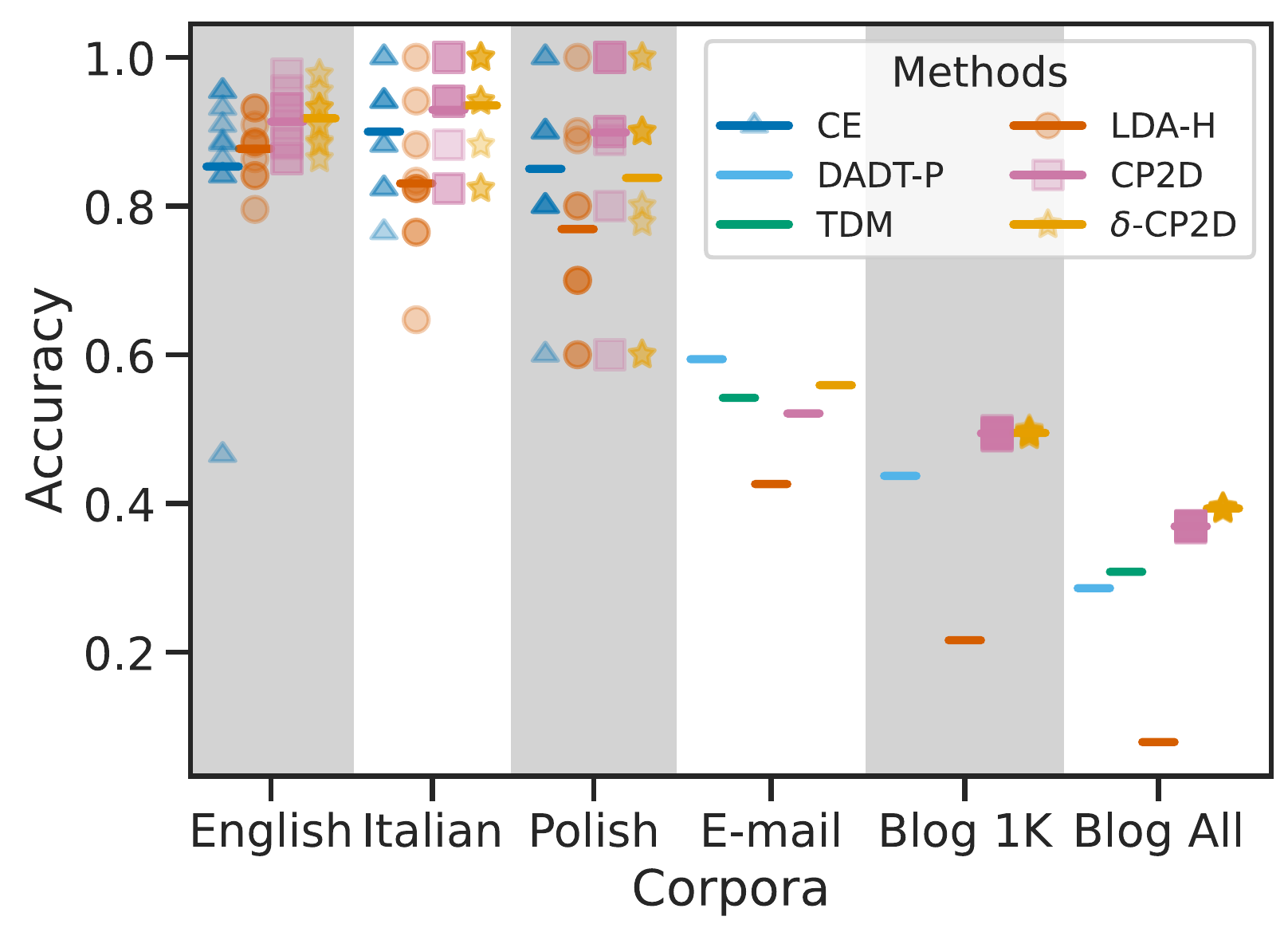}
\caption{{ \bf Attribution accuracy}. 
For each of the considered datasets and attribution methods, thick lines show the average
accuracy in the ten-fold stratified cross-validation experiment, while shaded circles refer to the attribution accuracy  on  
 each of the ten test sets separately. 
 An exception is the E-mail dataset, where a unique test set is considered (see main text). 
We compare the accuracy achieved by our method (the Constrained Probability 2-parameters Poisson-Dirichlet, in both its versions with and without including the parameter $\delta$: the CP2D and $\delta$-CP2D) with the  Cross-Entropy based approach (CE), the Latent Dirichlet Allocation plus Hellinger distance (LDA-H), the Disjoint Author-Document Topic model in its Probabilistic formulation (DADT-P), and the Topic Drift Model (TDM).
 On the literary corpora, the LDA-H accuracy is computed using our implementation; please refer to Supplementary Methods for details. For the informal corpora, the results are are available from a previous study~\cite[Table 1]{seroussi2012authorship}.
Results for the DADT-P and the TDM algorithms were available in the  works by Seroussi et al.~\cite[Tables 4, 5]{seroussi2014authorship} and Yang et al.~\cite[Table 1]{yang2017authorship}, respectively.  }
 \label{fig:results}
\end{figure}

\subsection{Results}\label{sec:results}
We test our approach on literary corpora and informal corpora. To challenge the generality of our method versus language variation~\cite{rybicki2011deeper}, we consider three corpora of literary texts in three different languages, English, Italian, and Polish, belonging to distinct Indoeuropean families and bearing a diverse degree of inflection (refer to the Supplementary  Note 1 for details).
We further consider informal corpora mainly composed of English texts. They are particularly challenging for the attribution task due to the strong unbalance in the number of samples per author and the texts' lengths (refer to Fig.~\ref{fig:datasets}, panel a).  
We consider, in particular, an email corpus and a blog corpus. The first is part of the Enron Email corpus proposed during the PAN'11 contest~\cite{argamon2011overview}. It is still used as a valuable benchmark, and we  compare the accuracy of our method with  those reported in~\cite{yang2017authorship,seroussi2014authorship}. The Blog corpus is one of the largest datasets used to test methods for authorship attribution~\cite{saedi2021siamese}. 
This is a collection of 678,161 blog posts by 19,320 authors taken from~\cite{schler2006effects}. Additionally, in line with~\cite{seroussi2011authorship,seroussi2012authorship}, we test our method on the subset of  1,000 most prolific authors of this corpus.
For more details on the corpora, please refer to the Supplementary  Note 1.

In Fig.~\ref{fig:datasets} (panels b--f), we illustrate the dependency of the attribution accuracy on the value of two free parameters of our model, specifically the normalisation $\delta$ and the length of the fragments in which we partition the text to be attributed.
In particular, we report the accuracy achieved on each dataset in a leave-one-out experiment, where we select each text in turn and attribute it by training the model on the rest of the corpus  (refer to the  Supplementary  Methods for more details).
 We note that, although simply setting $\delta = 1$ often gives the most or nearly the most accurate results, in a few datasets using a different value of $\delta$ significantly improves the accuracy. 
Indeed an effect of $\delta$ is also to correct for a non-optimal choice of the length of the fragments, as is evident in the literary English dataset.
When attributing an anonymous text, we optimise these two parameters---as well as the selection of $P_0$, the definition of the tokens, and the strategy to attribute the whole document from the likelihood of single fragments---on the training and validation sets, as detailed in the Supplementary  Methods.

In the case of informal corpora, we compare our method with state-of-the-art methods in the family of
topic models~\cite{Blei2010ProbabilisticTM}. Topic models are among the most established applications of nonparametric Bayesian techniques in natural language processing, and different authors' attribution methods rely on this approach. The underlying idea is to consider each document as a mixture of topics and to compute the similarity between two documents in terms of a measure of overlap between them (as detailed in the Methods section and the Supplementary  Methods).
Those methods were proposed to address challenging situations, particularly in informal corpora with many reference authors and typically short texts. Moreover, they have a similar ground to the method we propose. 
We consider the Latent Dirichlet Allocation plus Hellinger distance (LDA-H)~\cite{seroussi2011authorship},
the Disjoint Author-Document Topic model in its Probabilistic version (DADT-P)~\cite{seroussi2014authorship} and the Topic Drift Model (TDM)~\cite{yang2017authorship} since their performances are available on the informal corpora. LDA-H 
is a straightforward application of topic models to the authorship attribution task. The DADT-P  algorithm is a generalisation of the LDA-H  characterising both the topics associated with texts and with authors.
 TDM merges topic models with machine learning methods~\cite{yang2016discovering,mnih2013learning}  to account for dynamical correlations between words.

For the literary corpora, there is no direct comparison available in the literature. In the family of topic models, we considered the LDA-H approach,  whose implementation is available with the need for minor intervention (please refer to the Supplementary  Methods for details on our implementation). 
In addition, we consider a cross-entropy (CE) approach~\cite{PhysRevLett2002,baronchelli2005artificial}  in the implementation used 
 in previous research~\cite{lalli2018ferran}.
Compression-based methods are general and powerful tools to assess similarity between symbolic sequences
and have been at the forefront of authorship attribution for considerable time~\cite{neal2017surveying}.

When comparing the aforementioned methods and ours, we optimise the free parameters of our model (i.e., $\delta$, length of fragments, attribution criterion, type of tokens, and $P_0$)  on the training set,  as detailed in the Supplementary  Methods. 
The email corpus already provides training, validation, and test sets. For the remaining corpora, we 
use ten-fold stratified cross-validation~\cite{seroussi2011authorship,seroussi2012authorship,seroussi2014authorship}: 
 in turn, one-tenth of the dataset is treated as a test set and the other nine-tenths as training, and the number of samples per author is kept constant across the different folds.
In Fig.~\ref{fig:results}, we report the accuracy obtained on each of the ten partitions,  as well as the average value over them.  We show the results obtained by either switching off the parameter $\delta$ (that is, by fixing it to $1$) or optimising it on each specific corpus. The first scenario is denoted by CP2D (Constrained Probability 2-parameters Poisson-Dirichlet), the latter by $\delta$-CP2D.
The second procedure yields better performances in all the datasets
except for the Polish literary dataset, where the number of texts per author is too low to prevent overfitting in this simple training setting. 
In the literary corpora, the attribution accuracy is overall high, and that of our method consistently higher than that of the other techniques. 
In the informal corpora, our method achieves an accuracy slightly lower than the best-performing algorithm on the email corpus, while it is the most accurate on the blog corpus. 
This latter corpus presents a very large number of candidate authors, and our approach appeared more robust in these extreme conditions.
In Table~\ref{tab:all}, we present the numerical value of the average accuracy over the ten partitions, as shown in Fig.~\ref{fig:results} (additional evaluation metrics can be found in the Supplementary  Results). We also add the attribution accuracy on the training set. 
We observe that in the literary corpora, only in the Polish dataset, the accuracy on the test set is significantly lower than that in the training set, pointing to overfitting, as discussed above.
For the informal corpora, we conversely notice an increase in attribution rate from the training to the test corpora. For the email corpus, also other methods exhibit a similar behaviour~\cite{seroussi2012authorship,seroussi2014authorship}. This is probably related to the particular partition considered. For the Blog corpus, the attribution accuracy on the test set is not available for the other methods. 
Our method features a slightly greater accuracy on the test set than on the training, suggesting that, on the one hand, the corpus is sufficiently large to prevent overfitting. On the other hand, the method increases accuracy when increasing the length of the reference authors' sequences. 

\section{Conclusion}

We present a method for authorship attribution based on
urn models for innovation processes. We interpret texts as instances of stochastic processes, where the generative stochastic process represents the author. The attribution relies on the posterior probability of the anonymous text being generated by a particular author and continuing their production.
We consider the UMT model~\cite{tria2014dynamics} in its exchangeable version~\cite{tria2018zipf,tria2020}, which is equivalent to the two-parameter Poisson-Dirichlet process. While the latter process is widely used in Bayesian nonparametric inference, it is often employed in a hierarchical formulation. In the case of attribution tasks, this approach has led to topic models, where the output of the stochastic process is a sequence of topics, i.e., distribution over words. Here, we follow a more direct approach, where the stochastic process directly generates words. By relying on a heuristic approach, we can explicitly write posterior probabilities that can be computed exactly. Besides its computational convenience, the method we propose is easily adaptable to incorporate more realistic models for innovation processes. 

For instance, one avenue we intend to explore in future research is leveraging the urn model with semantic triggering~\cite{tria2014dynamics}.

We evaluate the performance of our approach by employing the simple UMT exchangeable model against various related approaches in the field. Specifically, we compare it with information theory-based methods~\cite{PhysRevLett2002,baronchelli2005artificial,lalli2018ferran} and probabilistic methods based on topic models~\cite{seroussi2014authorship,yang2017authorship}. Our method achieves overall better or comparable performance 
in datasets with diverse characteristics, ranging from literary texts in different languages to informal texts.

We acknowledge that our method may not compete with deep learning-based models (DL) when large pre-training datasets are available~\cite{Fabien2020Bertaa,Bauersfeld2023}. 
Nonetheless,  it exhibits  robustness in challenging situations for DL, for example when only a few texts are available for many authors~\cite{Bauersfeld2023} or in languages where pre-training is less extensive~\cite{SVNrussian2021}.
A deeper comparison with deep learning-based approaches, perhaps by concurrently exploring more sophisticated urn models in our approach, is in order but beyond the scope of the present work (refer to the Supplementary  Results for a more detailed discussion and a preliminary analysis).

As a final remark, we also note that we have here considered the so-called closed-set attribution~\cite{Fadel2020overview}, where the training set contains part of the production of the author of the anonymous text. In open-set attribution~\cite{kestemont2019working,stamatatos2023overview},  the anonymous text may be of an author for which no other samples are available in the dataset.  Despite the conceptual differences and nuances between the two tasks, approaches based on closed-set attribution~\cite{stamatatos2023overview} are sometimes used also in open-set problems, for instance, by assigning the text to an unknown author if a measure of confidence falls below a given threshold. Similar strategies can be employed with our method by leveraging the conditional probabilities of documents.

We finally note that the method presented here is highly general and can be valuable beyond authorship attribution tasks. Although we expect it to be particularly suitable when elements take values from an open set and follow an empirical distribution close to that produced by the model, it can be applied to assess the similarity between any class of symbolic sequences. 

\section{Methods}

\subsection{UMT and PD processes}\label{sec:UMT_PD}

In~\cite{tria2014dynamics}, a family of   urn models  with infinitely many colours was proposed to reproduce shared statistical properties observed in real-world systems featuring innovations. 
In this context, a realisation of the process is a sequence $x^t =x_1, \dots, x_t$  of extractions of coloured balls, where $x_{t}$ is the colour of the element drawn at time $t$, and the space of  colours available at a given time -- the urn -- represents the adjacent possible space.
The urn model with triggering (UMT)~\cite{tria2014dynamics} (and in a more general setting in~\cite{tria2018zipf,tria2020})
operates as follows:
the system evolves by drawing items from an urn initially containing a finite number  $N_0$ of balls of distinct colours. At each time step $t$, a ball is randomly selected from the urn, its colour registered into the sequence, and returned to the urn. 
If the colour of the drawn ball is not in the sequence $x^{t}$,  $\tilde{\rho}$ balls of the same colour and $\nu+1$ balls of entirely new colours, i.e., not yet present in the urn, are added to the urn. 
Thus, the occurrence of new events facilitates others by enlarging the set of potential novelties.
Conversely, if the colour of the drawn ball already exists in $x^{t}$, $\rho$ balls of the same colour are added to the urn.
Given the history of extractions $ x^{t}$,  the probabilities $b_t$ and $q_{c,t}$ that the drawing at time $t$  results in a new colour or  yields a  colour $c$ already present in $ x^t$ are easily specified for this model: 
\small
\begin{equation}\label{eq:UMT}
    \begin{split}
        &b_t = \frac{N_0 +  \nu D_t}{N_0+\rho t + a D_t} \\
       & q_{c,t} = \frac{\rho n_{c,t} + a -\nu}{N_0+\rho t + a D_t}
    \end{split}
\end{equation}
\normalsize
where $D_t$ and  $n_{c,t}$ are the number of distinct colours and the number of extractions of colour $c$ in the sequence $x^t$, respectively, and $a=\tilde{\rho} - \rho +\nu+1$. Different choices of the parameters $(\rho, \tilde{\rho}, \nu)$ lead to different scenarios, enabling the UMT model to capture the empirical properties summarised by Heap's, Zipf's and Taylor's laws. 
In the original formulation~\cite{tria2014dynamics}, only two values for  
 the parameter $\tilde{\rho}$ were discussed: $\tilde{\rho} = \rho$ or $\tilde{\rho}=0$;
the special setting $\tilde{\rho} = \rho - (\nu +1)$, which makes the model exchangeable, was later pointed out~\cite{tria2018zipf}. 
We remind that exchangeability refers to the property that the probability of drawing any sequence $x^t \equiv x_1,\dots,x_t$ of any finite length $t$ does not depend on the order in which the elements occur: $P(x_1,\dots,x_t)=P(x_{\pi(1)},\dots,x_{\pi(t)})$ for each permutation $\pi$ and each sequence length $t$.
In this case, upon a proper redefinition of the parameters, namely  $ \nu/\rho \equiv \alpha $ and $N_0/\rho \equiv \theta$, 
the UMT model reproduces the conditional probabilities associated with the PD process (expressed in Eqs.~\eqref{eq:pPD}). We note here that such probabilities include the Dirichlet process as a special case, where $\alpha =0$ and $D_t$ grows logarithmically with $t$. In the framework of urn models, the Dirichlet process finds its counterpart in the Hoppe model~\cite{Hoppe_1984} and in the exchangeable version of the UMT model with the additional choice  $\nu=0$.
The PD process is defined by $0< \alpha < 1$ and predicts the asymptotic behaviour  $D_t \sim t^{\alpha}$~\cite{pitman2006combinatorial}.
We note that the probabilities in Eqs.~\eqref{eq:UMT} coincide, when renaming the parameters as stated above,  with those in  Eqs.~\eqref{eq:pPD}.

\subsection{The strategy  for $P_0$}\label{ssec:discreteP}

When $P_0$ is a discrete probability distribution (it has atoms), an already seen value $y$ can be drawn again from it, and the conditional probabilities no longer have the simple form as in Eqs.~\eqref{eq:pPD}. 
In this case, the conditional probabilities depend not only on the sequence $x^t$ of observable values but also on 
latent variables indicating, for each element in $x^t$, whether it has been drawn from $P_0$ or arose from the reinforcement process~\cite{buntine2010bayesian}.
In particular, we can define,  for each type $y_i$ ($i=1,\dots,D_t$) in $x^t$, a latent variable $\lambda_{i,t}$ that counts 
 the number of times $y_i$  is drawn from the base distribution $P_0$. The  probabilities  conditioned on the  observable sequence $x^t$ and on the latent variables sequence $\lambda^{D_t}$ read:
\small
 \begin{equation}\label{eq:pPD_discr}
    \begin{split}
P(x_{t+1}= y  \mid x^t,\lambda^{D_t}) =  \frac{n_{y,t}\!-\lambda_{i,t}\alpha}{\theta\!+\!t} + \frac{\theta\!+\!\alpha \Lambda_t}{\theta\!+\!t}P_0(y) &\\
 \text{if} \,\,\, n_{y,t}>0 &\\
P(x_{t+1}= y  \mid x^t,\lambda^{D_t}) =  \frac{\theta\!+\!\alpha \Lambda_t}{\theta\!+\!t}P_0(y )\;  \; \;\;\;\; \text{if} \,\,\, n_{y,t}=0&
\end{split}
\end{equation}
\normalsize
Where $\Lambda_t \equiv \sum_i^{D_t} \lambda_{i,t}$ is the total number of extractions from $P_0$ till time $t$. To compute the probabilities conditioned to the observable sequence $x^t$, we must integrate out the latent variables. This is an exponentially hard problem and efficient sampling algorithms~\cite{Blei2010ProbabilisticTM,teh2006collapsed,porteous2008fast} have been developed for an approximate solution.

By taking the perspective of the urn model, we investigate the possibility of bypassing the problem by imposing that each element can be extracted only once from $P_0(\cdot)$, which is equivalent to fixing all the latent variables $\lambda_{i,t}=1$ and set to zero the last term in the first equation in Eqs.~\eqref{eq:pPD_discr}.

The latter procedure effectively replaces   $P_0(y) $ with a history-dependent probability, normalised at each time over all the elements $y$ not already appeared in $x^t$. It reads:
\begin{equation}\label{eq:P0_history}
 P_0^{t}(y) \equiv P_0(y \mid \! y \notin x^t)= \begin{cases} \frac{P_0(y)}{1-\sum\limits_{\tilde{y}\in x^t} P_0(\tilde{y})}
      & \text{if } y \notin  x^t \\
      0 & \text{otherwise}
    \end{cases}
 \end{equation}
where the sum is over all the elements already drawn at time $t$. Note that this choice
breaks the exchangeability of the process with respect to the order in which novel elements are introduced.
In the implementation of our algorithm, we follow an even simpler and fast procedure, which yielded equivalent results. We simply introduce an author-dependent base distribution by considering, for each author $A$, the frequency of the tokens that do not appear in $A$. 
Such procedure translates into replacing  $P_0(y) $ with  $P_0^{(A)}(y) = \frac{P_0(y)}{P(A^{\mathcal{C}})}$,
where $A^{\mathcal{C}}$ denotes the set of all distinct tokens that do not appear in $A $.
This author-dependent base distribution proved to be preferable to simply using the original frequency, especially in datasets with short texts and few samples for each author.

\subsection{LDA and topic models}
LDA is a generative probabilistic model~\cite{Blei2003}, which generates corpora of documents.
A document is a finite sequence of words $w_1,w_2,...,w_N$ and it is represented as a random mixture over latent topics. Each topic corresponds to a 
categorical probability distribution over the set of all possible words.
Topics can be shared by different documents.
The total number $k$ of topics is fixed a priori and to each topic $i$ in each document $d$ is associated a probability $\theta_{i,d}$,
extracted independently for each document from a $k$-dimensional Dirichlet distribution $D(\alpha_1,\dots,\alpha_k)$.
 Each document $d$ is generated as follows: first, its length $N_d$  is extracted from a Poissonian distribution with a given mean. Then, the document is populated with words using the following procedure: a topic $i$ is extracted with probability $\theta_{i,d}$ and a word $w$ is extracted from $i$ with the probability associated to it in topic $i$. The probabilities $p_i(w)$ of a word $w$ in the topic $i$ is in turn extracted independently from 
a $W$-dimensional Dirichlet distribution $D(\beta_1,\dots,\beta_W)$, where $W$ is the total number of words $W$ in the corpus.

As in Eqs.~\eqref{eq:pPD_discr}, we can introduce latent variables~\cite{Blei2003}, now with a different meaning. To each word $w_{i,d}$ in document $d$, $i=1,\dots, N_d$, we associate a latent variable
 $\lambda_{i,d}$ that is the identifier of the topic $j$ from which the word $w_{i,d}$ is extracted.
 The joint distribution of the sequence of words $w^{N_d} \equiv w_{1,d} ,\dots, w_{N_d,d}$ and latent variables $\lambda^{N_d} \equiv \lambda_{1,d} ,\dots, \lambda_{N_d,d}$ in a document $d$ thus read:
\begin{equation}
P(w^{N_d} ,\lambda^{N_d} ) =
 \prod_{n=1}^{N_d} p(w_{i,d} |\lambda_{1,d})p(\lambda_{i,d} )
 \end{equation}
where $p(\lambda_{i,d} )\equiv \theta_{i,d}$.
To compute the posterior probability of the observable sequence $w^{N_d}$ we must integrate out the latent variables. This is an exponentially hard problem and is solved with methods for numerical approximation by using for instance Markov Chain Monte Carlo algorithms.
A more flexible approach is to use the Dirichlet or PD processes instead of the Dirichlet distributions over topics. 
This allows the number of topics $k$ to remain unspecified \textit{a priori}. 

The probabilities $\theta_{i,d}$ are the elements of a sequence generated by a Dirichlet or PD process, for each document $d$. The processes characterising each document share the same discrete base distribution, which is, in turn, generated by a Dirichlet or PD process with a non-atomic $P_0$. 
Again, efficient sampling algorithms for computing the posterior distributions~\cite{Blei2010ProbabilisticTM,teh2006collapsed,porteous2008fast} have been developed in this framework.

In the framework of authorship attribution, methods relying on LDA are more widely adopted than those based on the Dirichlet or PD processes, primarily due to their simplicity and comparable accuracy~\cite{seroussi2014authorship}.

The procedure followed by the LDA-H algorithm to address the author attribution task is described in the Supplementary  Methods.

\section{Data Availability}
The corpora used to validate our approach, with the exception of the Italian literature, are available online at the following addresses: Blog corpus (\url{https://u.cs.biu.ac.il/~koppel/BlogCorpus.htm}), PAN’11 email corpus (\url{https://doi.org/10.5281/zenodo.3713246}), Polish literature (\url{https://github.com/computationalstylistics/100_polish_novels}), English literature (\url{https://github.com/GiulioTani/InnovationProcessesInference/tree/main/sample_data/English_literature}). The Italian literature corpus is currently covered by copyright, we make accessible the list of included titles (\url{https://github.com/GiulioTani/InnovationProcessesInference/tree/main/sample_data/Italian_literature}). Please refer to the Supplementary Note 1 for details about which parts of the public datasets where used in this study.

\section{Code Availability}
All the code we used to compute all the attributions with the CP2D approach is publicly available under the GNU GPL v3.0 license at \url{https://github.com/GiulioTani/InnovationProcessesInference}.

\section{Acknowledgements}
M.L. acknowledges support by the European Project: XAI: Science and technology for the eXplanation of AI decision making
(https://xai-project.eu/), ERC-2018-ADG G.A. 834756. G.T.R. was supported by the Czech Science Foundation project No. 21-17211S.
This work has been realised in the framework of the agreement between Sapienza University of Rome and the Sony Computer Science Laboratories.

\section{Author contributions}
F.T. conceived and designed research;  F.T. and M.L. performed a preliminary investigation, collected in M.L. master thesis. M.L. and G.T.R. collected the datasets. G.T.R. wrote the code and performed the numerical calculations.   F.T. and G.T.R. analyzed the data, refined the model and discussed the results. All authors contributed to write the paper and revise the manuscript.

\section{Competing interests}
The authors declare no competing interests.

\section{Additional information}
{\bf Supplementary Information } is attached to this manuscript.

\bibliography{Main_natcom}

\begin{table}[ht]
    \centering
    
    \begin{tabular}{lcccccc}
        \toprule
          &Eng &Ita & Pol & Email & Blog1K &Blog \\
        \midrule
        LDA-H &  .877\footnotemark[1] &  .830\footnotemark[1] &  .769\footnotemark[1] &.426 & .216 &.079 \\
        CE &.853 &.900 &.870 & --- &  --- & --- \\
        DADT-P& --- & --- & --- &  \textbf{.594} & .437 &.286 \\
        TDM& --- & --- & --- &.542 &  --- &.308 \\
        CP2D &.913 &.929 &  \textbf{.899} &.521 &  \textbf{.494} &  \textbf{.369} \\
        $\delta$-CP2D&\textbf{.918} &  \textbf{.935} &.838 &.558 &\textbf{.495} &  \textbf{.393} \\
        CP2D${}^{\textsc{tr}}$  &.924 &.927 &.949 &.497 & .489 &.358 \\
        $\delta$-CP2D${}^{\textsc{tr}}$ &.926 &.929 &.956 &.518 & .490 &.386 \\
        \bottomrule
         \multicolumn{2}{l}{\footnotemark[1] \tiny Our implementation.}&&&&&\\
    \end{tabular}
    \caption{{\bf Attribution results}. 
 Numerical values of the average accuracy, depicted in Fig.~\ref{fig:results}, are here listed for all considered methods: our Constrained Probability 2-parameters Poisson-Dirichlet, in both its versions with and without including the parameter $\delta$ (the CP2D and $\delta$-CP2D), the  Cross-Entropy based approach (CE), the Latent Dirichlet Allocation plus Hellinger distance (LDA-H), the Disjoint Author-Document Topic model in its Probabilistic formulation (DADT-P), and the Topic Drift Model (TDM).   
    In addition, we report the average accuracy obtained on the training sets (TR) by our method, as discussed in the text.  We highlight in bold the best performance on each dataset.}
 \label{tab:all}
\end{table}

\end{document}


\preprint{}
\renewcommand{\theequation}{S\arabic{equation}}
\renewcommand{\thetable}{S\arabic{table}}
\renewcommand{\thefigure}{S\arabic{figure}}
\renewcommand{\thesection}{S.\Roman{section}}

\title{Supplementary Information to the Manuscript ``Inference through innovation processes tested in the authorship attribution task''}

\author{Giulio Tani Raffaelli}
\affiliation{%
 Institute of Computer Science of the Czech Academy of Sciences, Pod Vodárenskou věží 2, 182 07 Prague, Czech Republic
}%
\author{Margherita Lalli}
\affiliation{
Scuola Normale Superiore, Piazza dei Cavalieri 7, Pisa, Italy
}%
\author{Francesca Tria}
\affiliation{%
 Sapienza University of Rome, Physics Department, Piazzale Aldo Moro 5, 00185 Rome, Italy
}%
\maketitle

\section{Supplementary Note 1: Corpora}
\label{sec:corpora}
In the main text, we considered different corpora to test our approach under diverse conditions. We chose literary corpora to investigate the effect of different languages while using long texts from a few tens of authors. On the other hand, we used two English corpora of informal texts -- one of email, the other of blog posts -- to evaluate our approach in more challenging cases with thousands or hundreds of thousands of shorter texts.

\paragraph{The Literary Corpora.} 
Using different languages may present challenges for Authorship Attribution. For example, in changing language, a word may have a single form or many, depending on its semantic role.
Adjectives have one form in English (\textit{red}), up to four in Italian (\textit{rosso}, \textit{rossa}, \textit{rossi}, \textit{rosse}), ten in Polish (\textit{czerwony}, \textit{czerwona}, \textit{czerwone}, \textit{czerwonego}, \textit{czerwonej}, \textit{czerwonemu}, \textit{czerwoną}, \textit{czerwonym}, \textit{czerwonych}, \textit{czerwonymi}). 
This may introduce differences in favour of one or another language as investigated in the works of Eder and collaborators~\cite{eder2011style, rybicki2011deeper, eder2012birds, eder2015does}. 

To investigate this, we considered three corpora of literary texts: (1) English, (2) Italian, (3) Polish.

The English corpus consists of 439 books from 44 authors active in the UK around the end of the \nth{19} century. It is derived from the one used in~\cite{lahiri2013complexity}, from which we selected only native British English speakers from the UK alive in 1894 to ensure a relative linguistic uniformity.
Further, we adopt the following procedure to disentangle the effect of language variation from that of imbalance in the authors' representation (that will be investigated on informal corpora). For each author with more than ten books in the corpus, we only retain the ten books with length closest to the average length of her texts, plus one-tenth of the remaining production randomly sampled.

The Italian corpus contains 171 books from 39 contemporary Italian authors. It is derived from the one used during the workshop ``Drawing Elena Ferrante's Profile'' held in Padova, Italy in September 2017~\cite{Arjuna_Ferrante,Arjuna_workshop} extended as in~\cite{lalli2018ferran} but without the writings from Elena Ferrante. The author behind this \textit{nom de plume} was never officially identified, and the inclusion of her books would have added a significant source of noise.

The last of the literary corpora contains 100 books from 34 Polish authors. It is derived from a benchmark corpus of 100 Polish novels, covering the 19th and the beginning of the 20th century. We removed one book corresponding to the only instance by Magdalena Samozwaniec. Thus, we used 99 novels by 33 authors (1/3 female writers, 2/3 male writers). The Computational Stylistics Group at the Institute of Polish Language~\footnote{\url{https://github.com/computationalstylistics/100_polish_novels}} prepared the corpus for stylometric analysis.

\paragraph{The Informal Corpora.}  The Email and Blog corpora have  already been used as testbeds for other approaches and present  partly complementary features: The Email corpus is mainly composed of very short texts, while the Blog corpus contains a considerable number of authors. We detail the two corpora here below.

The Email corpus is one of the corpora used during the PAN'11 contest~\cite{argamon2011overview}. 
The corpus has been developed based on the Enron Email corpus~\footnote{\url{http://www.cs.cmu.edu/enron/}} and divided into sections to account for different  attribution and verification scenarios. 
We selected only one pair of the twelve separate training and test collections (of which six are dedicated to author verification). We use the ``Large'' training sets provided for authorship attribution containing 9337 documents by 72 different authors. As a test set, we chose the one containing only texts written  by  authors already present in the training set.
The tasks at PAN'11 intended to reflect a ``natural'' task environment, so the curators included texts in languages different from English or automatically generated.  The curators automatically replaced personal names and email addresses with <NAME/> and <EMAIL/> tags, respectively, and we maintain this procedure. 
Under any other respect, the curators declare that the texts are guaranteed to be identical to the original. The original curators of the corpus had determined authorship based on the ``\texttt{From}'' email headers. The curators tried to match all the addresses belonging to the same individual; however, they acknowledged some errors.

The last and largest dataset we considered is the Blog corpus. This is a collection of 678,161 blog posts by 19,320 authors taken from~\cite{schler2006effects}\footnote{Available for download from \url{u.cs.biu.ac.il/~koppel}}. This corpus is valued for its size and, even if it has been around for quite a long time, it is still in use to test novel stylometric approaches~\cite{saedi2021siamese, li2023knowledge,jeyaraj2023author}.
This corpus is challenging for the need to
 handle many authors/texts on one hand, and for the presence of texts of very short length on the other hand ($\sim$0.5\% of the posts have at most ten characters, and $\sim$11.5\% have no more than 100).
Many authorship attribution approaches  require the corpus to be reduced~\cite{jafariakinabad2021unifying,patchala2018authorship,saedi2021siamese}. 
For this reason, in addition to the whole corpus, we also analyse  a subset of the 1000 most prolific authors to compare with results in~\cite{seroussi2011authorship, seroussi2012authorship}.

\paragraph*{} Every text is encoded by using a single-byte encoding ISO 8859-1 for Italian and English texts, ISO 8859-2 for the Polish ones. While this forces transliterating some rare off-encoding characters (e.g. a Greek quotation in an English book), it simplifies feature extraction.  From every corpus, we excluded authors that had only one book.

\section{Supplementary Methods}
\subsection{Estimate of discount and concentration parameters}
The conditional probability $P(T \mid \!A )$ of a  text given an author---as defined in the main text---depends on  the discount and concentration parameters ($\alpha$ and $\theta$ respectively) of the author's process. 
These are obtained for each author by maximizing the probability (likelihood) of  the sequence defined by the concatenation of their texts.
From Eq.~(1) (in the main text) we can derive the joint probability of an author sequence $x^m \equiv x_1, \dots, x_m$:
\begin{equation}
    P(x_1,\ldots,x_m \mid \alpha,\theta,P_0)=P(n_1,\ldots,n_k\mid \alpha,\theta,P_0)= \frac{\left(\theta \mid \alpha \right)_{k}}{\left(\theta\right)_n}\prod_{j=1}^{k} \left[ P_0(y_j) \left(1-\alpha \right)_{n_j-1}\right]
    \label{eq:seq}
\end{equation}
where we follow the notation established in the main text and $k$ is the number of distinct tokens in $x^m$.
We now note that the values of $\alpha$ and $\theta$ that maximise~\eqref{eq:seq} do not depend on the base distribution $P_0$,
 so that they can  be estimated by only considering the probability  associated with the partition of the sequence $x^m$ in $k$ classes:
 
\begin{equation}
    (\alpha_A,\theta_A)=\arg \max_{(\alpha,\theta)} \frac{\left(\theta \mid \alpha \right)_{k}}{\left(\theta \right)_n}\prod_{j=1}^{k} \left(1-\alpha \right)_{n_j-1}
    \label{eq:argmax}
\end{equation}

  We search for the maximum by relying on the  gradient ascent with momentum.
  In order to do this, we first need to evaluate the derivatives of the probability in~\eqref{eq:argmax} with respect to $\alpha$ and $\theta$. 
Note that from the moment they will have to be set equal to zero, a proportional quantity will serve as well. 
Then, expressing the Pochhammer symbol with the $\Gamma$ functions, we have:
\begin{eqnarray}
\frac{\partial P}{\partial \theta}&\propto& \frac{\partial}{\partial \theta}\left[ \alpha^{k} \frac{\Gamma\left(\frac{\theta}{\alpha}+{k}\right)}{\Gamma\left(\frac{\theta}{\alpha}\right)}\frac{\Gamma\left(\theta\right)}{\Gamma\left(\theta+n\right)}\right] \nonumber\\
&=&a^{k}\times \left[ 
        \frac{1}{\alpha}\frac{\Gamma'\left(\frac{\theta}{\alpha}+{k}\right)}{\Gamma\left(\frac{\theta}{\alpha}+{k}\right)}
        -\frac{1}{\alpha}\frac{\Gamma'\left(\frac{\theta}{\alpha}\right)}{\Gamma\left(\frac{\theta}{\alpha}\right)}
        +\frac{\Gamma'\left(\theta\right)}{\Gamma\left(\theta\right)}
        -\frac{\Gamma'\left(\theta+n\right)}{\Gamma\left(\theta+n\right)}
    \right] \times \frac{\Gamma\left(\frac{\theta}{\alpha}+{k}\right)}{\Gamma\left(\frac{\theta}{\alpha}\right)}\frac{\Gamma\left(\theta\right)}{\Gamma\left(\theta+n\right)}  \nonumber \\
     &\propto& \frac{1}{\alpha} \left[
        \psi^0 \left(\frac{\theta}{\alpha}+{k}\right) - \psi^0 \left(\frac{\theta}{\alpha}\right)
    \right] + \psi^0 \left(\theta\right) - \psi^0 \left(\theta+n\right) \nonumber\\
    &\equiv &a_{\theta}(\alpha, \theta \mid n,{k}), \label{eq:detheta}\\
    \frac{\partial P}{\partial \alpha} &\propto& \frac{\theta}{\alpha^2}\left[
        \psi^0 \left(\frac{\theta}{\alpha}\right)-\psi^0 \left(\frac{\theta}{\alpha}+{k}\right)
    \right]+\frac{{k}}{\alpha}+\sum_i r_i \left[
        \psi^0 \left(i - \alpha\right)-\psi^0 \left(1-\alpha\right)
    \right] \nonumber\\
    &\equiv&  a_{\alpha}(\alpha, \theta \mid {k}, \textbf{r}).
    \label{eq:deralpha}
\end{eqnarray}
where  $\Gamma(z)$ is the Euler $\Gamma-$function and $\psi^0(z)$ is the Digamma function, defined as 
\begin{equation}
    \psi^0\left(z\right) \equiv \frac{\mathrm{d}}{\mathrm{d} z}\ln\Gamma\left(z\right) = \int_0^{\infty} \left( \frac{e^{-t}}{t} - \frac{e^{-zt}}{1-e^{-t}} \right) \mathrm{d} t;
\end{equation}
 we have introduced $r_i$ to indicate the number of tokens with multiplicity $i$, with $\sum_i r_i = {k}$ and $\sum_i i r_i = n$.

In the  gradient ascent algorithm with momentum,
we can regard $\alpha$ and $\theta$  as the coordinates of a unit mass point moving in a damping medium, i.e., subject to a  force proportional to the speed that opposes the movement~\cite{qian1999momentum}. 
The recursive equations  can be written as:   
\begin{equation}
    \begin{cases}
        \alpha(t\!+\!1)=\alpha(t)\! +\! I_{\alpha}v_{\alpha}(t\!+\!1)\\
        v_{\alpha}(t\!+\!1)=\eta v_{\alpha}(t)\!+\!a_{\alpha}(t)
    \end{cases}
    \begin{cases}
        \theta(t\!+\!1)=\theta(t)\!+\!I_{\theta}v_{\theta}(t\!+\!1)\\
        v_{\theta}(t\!+\!1)= \eta v_{\theta}(t)\!+\!a_{\theta}(t)
    \end{cases}\label{eq:descent}
\end{equation}
where  $1-\eta$  are dissipation terms.  The gradient ascent algorithm with momentum is known to increase the rate of convergence with respect to a simple gradient ascent, which is recovered in the case $\eta =1$. 

To improve and speed up the convergence around the maximum, we added a second feature. 
When a derivative changes sign, we consider  a step of the bisection method: we replace the value of $\alpha$ or $\theta$, depending on the derivative of whom changes sign,
 with its average over the last two steps, and we halve the corresponding value of the speed. 
 
 We also add a correction procedure that ensures obtaining admissible values for $\alpha$ and $\theta$: when they  exit from their domain ($\alpha \in [0,1)$ and  $\theta > -\alpha$), they are reset to 
 a value close to the border (0.01 and 0.99 for $\alpha$ and $-\alpha+0.1$ for $\theta$), and we continue the optimisation from this new starting point. Each time we need to adopt this correction procedure, we modify the parameters in~\eqref{eq:descent} by halving $\eta$ and 
 the scaling $I_s$ ($s=\alpha,\theta$) of the variable that exited the admissible range.
If we need to adopt this correction procedure more than $50$ times, we exit the optimisation and fix the values of  $\alpha$ and $\theta$ to their current values or to a value close to the border (0.001 and 0.999 for $\alpha$ and $-\alpha+0.01$ for $\theta$) if the current value is outside the admissible range. 
 
The starting values we chose for $\alpha$ and $\theta$ are respectively 0.3 and $k$ (the number of distinct tokens), while we initially set  $I_\alpha=10^{-4}$, $I_\theta=1$ and $\eta =0.9$.
\subsection{LDA-H implementation}
The Latent Dirichlet Allocation (LDA)  is one of the most popular methods for topic modelling, firstly used for topic discovery in~\cite{blei2003latent}. 
It is based on the identification of  documents as random mixtures over latent topics, where each topic is characterised by a distribution over words.
The LDA-H is a variation of the LDA, proposed in~\cite{seroussi2011authorship}, that uses in addition the Hellinger distance to quantify the dissimilarity between  the stochastic  sources at the root of two different documents and in 2011 achieved state-of-the-art results in the field of authorship attribution.
The attribution scheme outlined  in~\cite{seroussi2011authorship} consists of two steps: in the first one, LDA is used to reduce the dimensionality of each document $i$ in the corpus to a topic distribution $\theta_i$, $i=1, \dots, T$, specifically a multinomial distribution of dimensionality $T$.
In the second step,
the  most likely author of a document is found by minimising the Hellinger distance between document topic distributions, which is defined as:
\begin{equation}
    D(\theta_1, \theta_2)=\sqrt{\frac{1}{2}\sum_{t=1}^T\left(\sqrt{\theta_{1,t}} - \sqrt{\theta_{2,t}}\right)^2}
\end{equation}
where $\theta_1$ and $\theta_2$ represent the topic distributions of the documents $1$ and $2$ and $\theta_{i,t}$ is the probability of the topic $t$ in document $i$. 

The authors propose two variants of the model~\cite{seroussi2011authorship}:
\begin{enumerate}
    \item Multi-document (LDAH-M). In this instance-based version, distances are computed between the topic distribution of the anonymous text and all the known texts. The author whose texts are on average closer to the anonymous one is returned as the most likely author of the test document.
    \item Single-document (LDAH-S). In this case, all the documents from each author are chained together in a profile document. The authorship of the anonymous text is attributed to the author whose profile document has the closest topic distribution.
\end{enumerate}
 Here we rely on the second approach, which was found to achieve better performance on large dataset~\cite{seroussi2011authorship}.
 The LDA uses as priors  for the distribution of topics  in each document  and for the distribution of words inside each topic 
   a $T$-dimensional symmetric Dirichlet distribution with parameters $\alpha$ and $\beta$ respectively.  The posterior distributions  are sampled through a collapsed Gibbs sampling. Once the distribution of words inside each topic is inferred using the test set,
   the topics distribution of  the anonymous text is  inferred. We  used the  \texttt{lda} module for Python~\cite{ldaModule} which in turn cites~\cite{blei2003latent, griffiths2004finding, Pritchard945} for its implementation. It differs from the original implementation from LingPipe~\cite{lingpipe}, which looks discontinued since 2011, for the method to estimate the topics distribution of the anonymous text, that relies on~\cite{buntine2009estimating, wallach2009evaluation}. Following~\cite{seroussi2012authorship}, we ran 4 chains with a burn-in of 1,000 iterations and, after the burn-in, we took 8 samples spaced 100 iterations. For each sample, we compute the Hellinger distance between the profile document and the  anonymous text topic distributions and then average over the 32 values obtained.
 
To fix the model parameters, we follow~\cite{seroussi2011authorship}, which in turn refers to~\cite{griffiths2004finding}, and we set
   $\alpha = \min (0.1, 50/T)$ and $\beta = 0.01$.  The number of topics is fixed to $T=100$, that is in the range of optimal values estimated for
  different datasets in~\cite{seroussi2011authorship}.

\subsection{Selection of parameter values}\setcurrentname{Selection of parameter values}\label{sec:parameter_selection}
The attribution process we presented depends on two parameters for each author ($\alpha$ and $\theta$) and five hyperparameters, namely the kind of tokens (e.g., dictionary words, OSF 9-grams), the strategy for $P_0$ (i.e., $P_0$ or $P_0^{(A)}$), the length of the fragments, the strategy for assigning texts from fragment probabilities (i.e., Maximum Likelihood or Majority Rule), and the value of $\delta$. As described in the main text, we use part of the database as a training set to select the set of hyperparameters.

The training sets of the literary corpora and the Blog corpus are defined by following the ten-fold stratified cross-validation scheme: ten times, one-tenth of the dataset is treated as a test set and the other nine-tenths as training. It is defined as ``stratified'' when the number of samples per author is kept constant across the different folds. This approach offers a good reference for the whole dataset and ensures that authors with very few texts always have some in the training set. Each of the ten training sets will bring different sets of hyperparameters and parameters (the corpus of each author will differ, and so will the estimate of $\alpha$ and $\theta$). This approach is more reliable than a simple 90-10 dataset split in training and testing, whose results can change dramatically with different splits (see the spread for the different folds in Fig.~2 in the main text). On the other hand, the email corpus has the split already provided with the dataset. We kept the same split for better comparison with other results in the literature and used as training set the union of the  train and validation sets reported in~\cite{argamon2011overview}.
 
 Once the test set is defined, we select the set of hyperparameters that maximises the accuracy on the training set.
 To do this, we use part of the training set as validation set. The definition of the latter depends on the dataset, in particular, we made different choices according to the corpus size.
 We applied the leave-one-out criterion for the literary corpora and the Email corpus. In this setting, each document of the training set, in turn, is treated as a validation set, and the author parameters are optimised on the rest of the training set. In the Blog corpus (substantially larger), we  partition  the training set in nine folds  and use, in turn, each of them as a validation set. We optimise in this case the average of the accuracy on the nine validation folds. Finally, the size of the complete Blog corpus brings on the one hand a significant computational burden and on the other hand very stable results. We thus optimise for $P_0$ and $\delta$ using a single ninth of the training set as validation. In addition, we do not optimise neither the token kind, which we fix to OSF 5-grams, nor the fragments' size, considering the whole document directly. This latter choice also makes  unnecessary the optimisation of the attribution criterion from the fragments probabilities.

 We search for the best hyperparameters on a grid, dividing the procedure into two steps for greater efficiency.
 We first optimise over the five hyperparameters, moving  $\delta$ in the range $\delta\in [10^{-2}, 10^1]$ and considering only four fragments' lengths. We then fix tokens and $P_0$ and refine the search for the best fragment length on 10 lengths: 9 logarithmically spaced from 1 to $10^4$ tokens, plus full documents.
  We  note that the correction $\delta$ to the normalisation of $P_0$ can be factored out by writing
\begin{equation}
    P(T\mid A,\delta) = P(T\mid A,1) \times \delta^{D^{T\cup A}-D^A},
    \label{eq:p_delta}
\end{equation}
which allows for an efficient evaluation of the probabilities with many different $\delta$ values. 

\begin{figure}
\centering
\includegraphics[width=\textwidth]{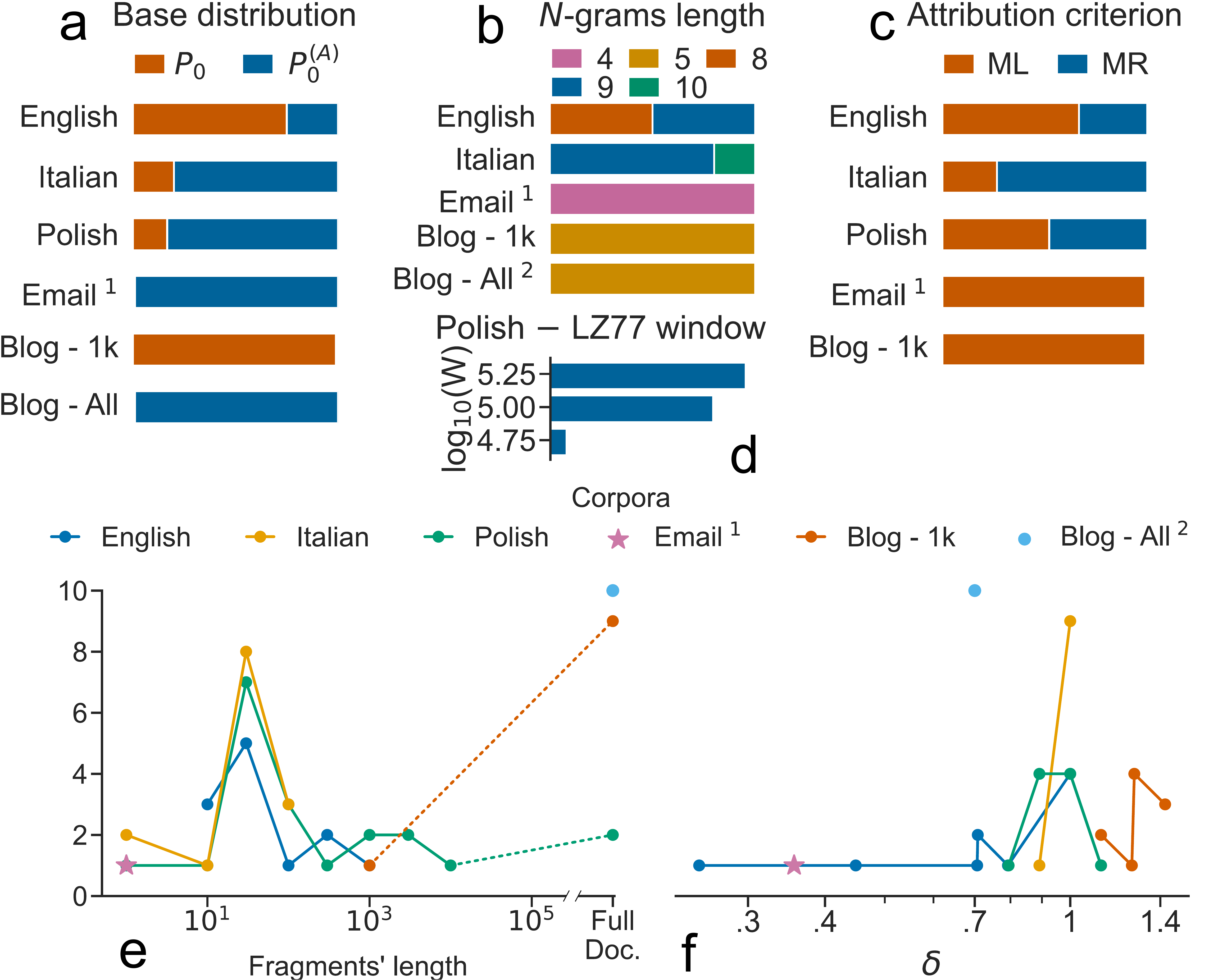}
\caption{\textbf{Optimal parameters for the attribution task.} We show the frequency of the various choices of parameters  that maximise the attribution on the training set, over the ten partitions and for each dataset. Panel (a): choice of the base distribution $P_0$ (see main text and above). Panels (b) and (d): kind of tokens. For each corpus either the $N$-grams or the LZ77 sequences  are always selected.  We achieve a higher accuracy with $N$-grams  for all the corpora with the exception of the Polish literary corpus, where the LZ77 sequences turned out to be the optimal choice. The optimal  $N$-grams length varies over different corpora, including $N=4$ (in purple), $N=5$ (in fuchsia), $N=8$ (in goldenrod), $N=9$ (in cyan), $N=10$ (in green), and only slightly inside each corpus. 
For the Polish literary corpus (panel (d)), we report a histogram of the logarithm of the best length of the LZ77 window (W). 
Panel (c): we compare the two considered  criteria for the attribution (see main text and above): Maximum Likelihood (in goldenrod) and Majority Rule (in cyan). 
In the bottom panel, we plot the number of folds in which a given length of fragments (panel (e)) and a given  value of the parameter $\delta$ (panel (f)) were optimal. Note that in panel (e) we can appreciate the fact that multiple values of the same parameters can lead to identical levels of accuracy. As a consequence, by summing up  the number of folds where different values of the fragment length are optimal, we can exceed the maximum $10$.
Notes: ${}^1$ the Email corpus has only one set of parameters as the test set is predefined. ${}^2$ we did not optimise the $N$-gram length and the fragment size for the whole Blog corpus (refer to the section above). We selected $N=5$ and whole document \textit{a priori}. We thus do not apply any attribution criterion from fragments.}
\label{fig:param}
\end{figure}

In Figure~\ref{fig:param}, we report the criteria and the hyperparameters' values that are optimal for each partition and dataset. Note that in the smaller corpora, different sets of parameters can lead to the same attribution accuracy on the training set. This effect is indeed relevant for the Polish corpus (where we find identical performance for 9 different values of the fragment length for one of the folds, up to 24 equivalent choices of parameters when all ten folds are considered), while is negligible in the Italian and English corpora (where 4 and 2 ties occur respectively) and is not observed in larger corpora.

\section{Supplementary Results}
\subsection{Effect of the tokens' choice}
\begin{figure}
\centering
\includegraphics[width=\textwidth]{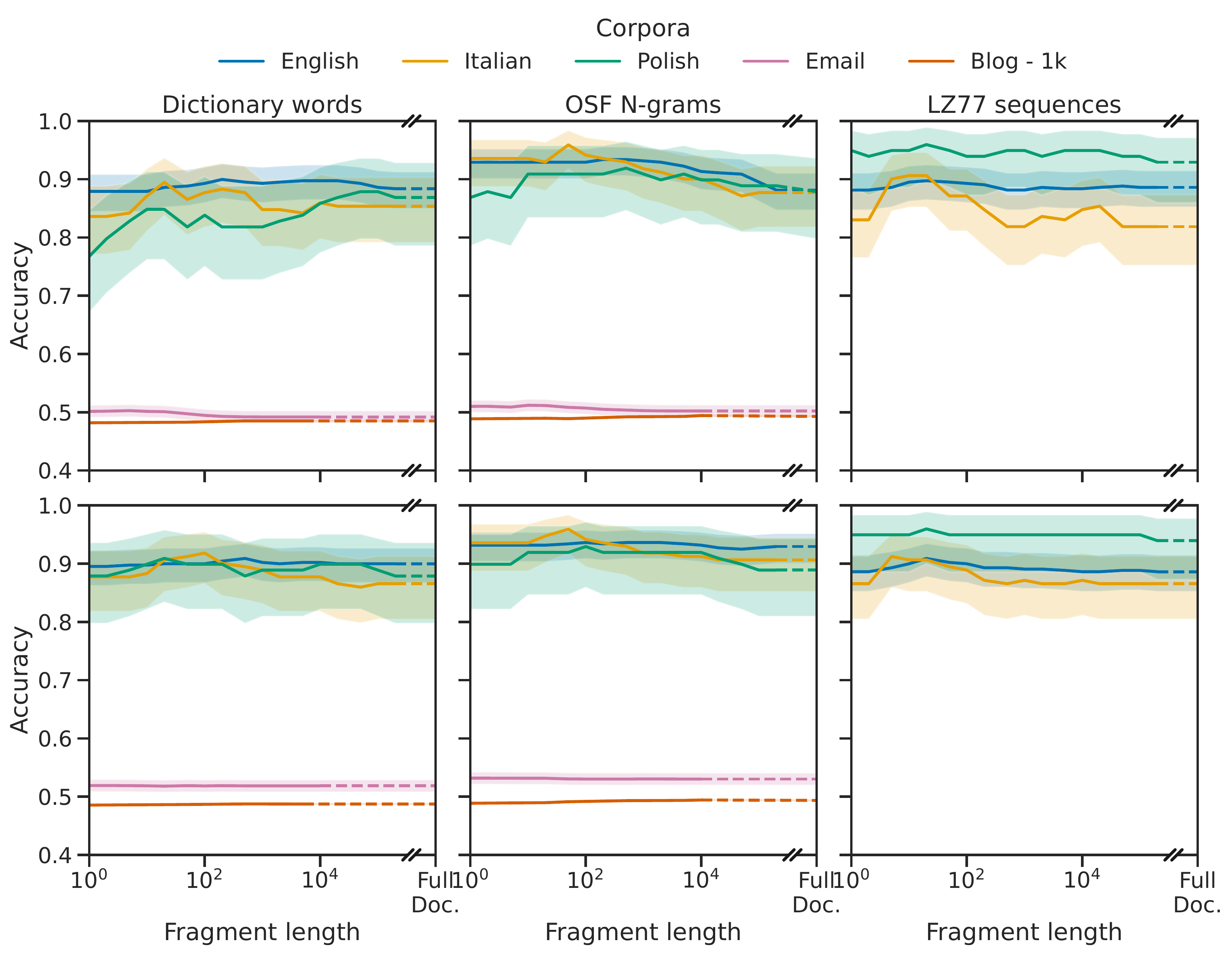}
\caption{\textbf{Results of a leave-one-out experiment for all corpora as function of fragments length.} We report the results for the accuracy using the different definitions of tokens:  dictionary words, Overlapping Space-free N-grams(\textsc{OSF} N-grams), and LZ77 sequences. In the top row the results with the global normalisation constant $\delta$ fixed to one. In the bottom row the results maximising over the values of $\delta$. These results are the maximum over the choices of $P_0$ and attribution rule. The shaded areas represent the 95\% confidence interval (Clopper–Pearson) on the accuracy assuming a binomial distribution of the correct attributions. Full Doc.: Full documents, no division into fragments.}
\label{fig:token}
\end{figure}

 In Figure~\ref{fig:token}, we show the attribution accuracy, for different choices of the tokens, as the length of the fragments varies. In all the plots, we maximised the attribution over the choices of the normalisation of $P_0$, the kind of attribution and the length $N$ of $N$-grams or the window size of LZ77. 
On the top panel, the value of $\delta$ equals 1, while in the bottom this is optimised as well.

We also include Dictionary Words obtained by removing punctuation and symbols and splitting the sequence using spaces. We do not apply any lemmatiser, and we acknowledge the presence of tokens like `doesen' that, even if it is not a regular English word, is the expected behaviour also of other tokenisation techniques, e.g.,~\cite{zhao2007entropy}.

We observe how one main effect of the optimisation over $\delta$ is to reduce the effect of the varying fragment length. Since the search on $\delta$ can be implemented in a very efficient way, the optimisation can compensate for a coarser search over the values of $F$.

\subsection{Comparison with Deep-Learning approaches}
Results of a deep learning-based approach (DL) for authorship attribution are available on subsets of the  Blog corpus analysed in the main text. It can thus be interesting to compare our results with those.

In Table \ref{tab:bertaa}, we contrast our results with those of  BertAA~\cite{Fabien2020Bertaa}.
BertAA relies on the Bert model, which is currently cutting edge in DL methods for authorship attribution~\cite{stamatatos2023overview,Bauersfeld2023}.
The authors in~\cite{Fabien2020Bertaa} consider the subsets of the 5, 10, 25, 50, 75 and 100 most prolific authors
in the Blog corpus. For each  dataset, they use 20\% of the texts as test set and the remaining as training set.
They use a stratified approach, meaning that they maintain the same proportion of number of texts for each author in the training and test set.
Accordingly, we 
consider a 5-fold stratified cross-validation experiment, i.e, we split the corpus into 5 folds and use in turn one of them as the test set, with a stratified approach. 

We can observe that BertAA achieves a higher accuracy than our method in this scenario. However, these results should be interpreted with caution. The accuracy of our method remains stable over the entire dataset (refer to the main text), where most  authors present a very limited production. There, no results for any DL-based model are available, to our knowledge. In addition, results reported in~\cite{Bauersfeld2023} for a method that refines BertAA suggest that the accuracy of DL approaches can significantly decrease when only a few samples per author are available in the training set.

More generally, DL approaches may be less effective when pre-training data is not extensive, for example, in languages other than English~\cite{Russian_DP2021}. As a final remark, it is undeniable that DL approaches are extremely powerful under suitable conditions; however, simpler and more intuitive approaches may be advantageous in complementary scenarios.

\begin{table}
    \centering
\caption{Comparison of the results using $\delta$-CP2D and BertAA~\cite{Fabien2020Bertaa} on subsets of the most prolific authors of the Blog corpus.}
\label{tab:bertaa}
    \begin{tabular}{llcccccc}
    \toprule
         \multicolumn{2}{r}{N-Authors:}&  5&  10&  25&  50&  75& 100\\
    \midrule
         \multirow{2}{*}{$\delta$-CP2D}&  Accuracy&  .517&  .479&  .514&  .467&  .469& .461\\
         &  Macro F1&  .427&  .433&  .515&  .455&  .467& .457\\
         \multirow{2}{*}{BertAA}&  Accuracy&  .613&  .654&  .653&  .597&  .609& .588\\
         &  Macro F1&  .491&  .588&  .640&  .564&  .583& .565\\
     \bottomrule
    \end{tabular}

\end{table}

\subsection{Additional measures of performance}

Accuracy and Macro-averaged F1 score are two of the most commonly used metrics for evaluating algorithm performance. Accuracy is defined as the proportion of items correctly classified. This measure is particularly suitable when the classes have a similar number of elements or when the cost associated with misclassifying an item does not depend on class abundance. Our case of authorship attribution follows in the second category.

We give here results in terms of the F1 score as well. 
The F1 score is given by the harmonic mean of Precision and Recall. 
Precision  measures the fraction of elements correctly assigned to a target class, among all those assigned to that class. Conversely, Recall measures
the fraction of elements, among all the elements belonging to a target class, that have been identified by the model.
For the sake of clarity, we report in Eq.~\eqref{eq:scores} the definition of Precision (P), Recall (R) and F1 score for a single class, i.e., author, $A$.
\begin{align}
    \text{P}_A =& \frac{\text{TP}_A}{\text{TP}_A+\text{FP}_A}\nonumber\\
    \text{R}_A =& \frac{\text{TP}_A}{\text{TP}_A+\text{FN}_A}\label{eq:scores}\\
    \text{F1}_A =& \:2\frac{\text{P}_A\cdot\text{R}_A}{\text{P}_A+\text{R}_A}\nonumber
\end{align}
where TP, FP, and FN are respectively True Positives, False Positives and False Negatives.
When considering the Macro-averaged F1 score, the F1 scores for each class are averaged with equal weight. 
This measure is more appropriate in applications where misattributing an instance from a less abundant class can have significant consequences.
In Table \ref{tab:acc_F1} we present the results for the F1 score obtained with the three algorithms we implemented. 

\begin{table}
    \centering
\caption{Accuracy and Macro-Averaged F1 scores for the three methods we implemented on the literary corpora.}
\label{tab:acc_F1}
    \begin{tabular}{clcccc}
    \toprule
         &  &  CP2D&  $\delta$-CP2D&  LDAH& CE\\
         \midrule
         \multirow{2}{*}{English}&  Accuracy&  .913&  \textbf{.918}&  .877& .853\\
         &  Macro F1&  \textbf{.892}&  .889&  .852& .830\\
         \multirow{2}{*}{Italian}&  Accuracy&  .929&  \textbf{.935}&  .830& .900\\
         &  Macro F1&  .925&  \textbf{.932}&  .813& .886\\
         \multirow{2}{*}{Polish}&  Accuracy&  \textbf{.899}&  .838&  .769& .870\\
         &  Macro F1&  \textbf{.887}&  .825&  .755& .843\\
         \bottomrule
    \end{tabular}

\end{table}

In addition, to allow a direct comparison with results reported in~\cite{argamon2011overview} for 
the competition PAN’11, we provide in Table~\ref{tab:PAN11}  the Micro- and Macro-averaged values of Precision, Recall, and F1 scores achieved by our algorithm in that dataset. 
We would have ranked sixth in the competition.  However, we note that  algorithms participating PAN'11 were most of the cases specifically optimised for this competition (e.g., including features specific to email messages such as opening greetings or closing formulas) or combined the results of multiple classifiers.

\begin{table}
    \centering
\caption{Micro- and Macro-averaged Precision, Recall and F1 scores on the Large test set from the PAN'11 competition~\cite{argamon2011overview}.}
\label{tab:PAN11}
    \begin{tabular}{lcccccc}
    \toprule
         &  \multicolumn{3}{c}{Micro-averaged}&  \multicolumn{3}{c}{Macro-averaged}\\
     \midrule
         &  Precision&  Recall&  F1&  Precision&  Recall& F1\\
         $\delta$-CD/DP&  \textbf{.558}&  \textbf{.558}&  \textbf{.558}&  \textbf{.482}&  .431& \textbf{.429}\\
         CP2D&   .521&  .521&   .521&  .473&  \textbf{.437}& .423\\
     \bottomrule
    \end{tabular}

\end{table}

\subsection{Computing times}

We here provide an estimate of the computing times on an ordinary laptop of the three algorithms we implemented. Table~\ref{tab:exec_times} reports the running time for one execution of a 10-fold cross-validated experiment performed on a laptop with 4 computing cores (2.8 GHz) and 16 GB of memory. These times are intended to guide the researcher interested in applying the code available at \url{https://github.com/GiulioTani/InnovationProcessesInference} to the corpora of their interest. The reported times include the entire procedure, from loading the texts to the output of the results.

Note that LDA-H and CE depend on hyperparameters (e.g., length of fragments for CE, number of topics for LDA-H) that we consider fixed in this analysis to the values suggested by the respective authors. For our approach, we report the total execution time both fixing the hyperparameters to given values and  including the 
full optimization procedure as described in section~\hyperref[sec:parameter_selection]{Selection of parameter values} in Supplementary Methods.
In the first case, we set the fragment length to $10$ tokens.
 In both cases, our approach appears much faster than CE or LDA-H.

To provide insight into how our method scales with the size of the training set, we also present in Table~\ref{tab:exec_blog} the running time for a 5-fold cross-validation experiment on  subsets of the Blog corpus.

\begin{table}[h]
    \centering
\caption{Running times for the three algorithms we implemented on the three literary corpora. All times are estimated on the same machine with 4 computing cores 2.80GHz and 16 GB of memory. The time format is hh:mm:ss.s.  }
\label{tab:exec_times}
    \begin{tabular}{lrlrcrc}
        \toprule
         &     \multicolumn{2}{c}{English}&\multicolumn{2}{c}{Italian}&  \multicolumn{2}{c}{Polish}\\
         \midrule
         CE&     6 days&11:23:07.5& &23:16:26.6&  & 18:59:26.0\\
         LDA-H&     3 days&23:16:18.9&1 day&10:22:37.8&  1 day& 03:00:29.7\\
         $\delta$-CP2D&     &00:03:06.7&&00:00:57.7&  & 00:05:32.2\\
         $\delta$-CP2D\footnotemark[1]&     &08:17:35.9&&03:48:58.0&  & 05:06:48.4\\
         \bottomrule
         \multicolumn{7}{l}{\footnotemark[1]  \footnotesize  Including optimisation of hyper-parameters as described in section}\\
         \multicolumn{7}{l}{\footnotesize{ \hyperref[sec:parameter_selection]{Selection of parameter values} in Supplementary Methods.}}\\
    \end{tabular}
\end{table}

\begin{table}[h]
    \centering
\caption{Running times of our algorithm on subsets of the Blog corpus. All times are reported on the same machine with 4 computing cores 2.80GHz and 16 GB of memory. Time format is hh:mm:ss.s.}
\label{tab:exec_blog}
    \begin{tabular}{lclrl}
    \toprule
         &  \multicolumn{2}{c}{$\delta$-CP2D}&  \multicolumn{2}{c}{$\delta$-CP2D\footnotemark[1]}\\
         \midrule
         Blog 5& \hspace{2em} &  00:00:09.3&  & 01:59:09.3\\
         Blog 10&  &  00:00:22.4&  & 03:38:05.4\\
         Blog 25&  &  00:00:54.3&  & 08:16:59.8\\
         Blog 50&  &  00:06:47.2&  & 15:33:01.3\\
         Blog 75&  &  00:08:57.7&  1 day& 02:25:57.9\\
         Blog 100&  &  00:11:09.1&  1 day& 08:27:11.4\\
         \bottomrule
         \multicolumn{5}{l}{\footnotemark[1]  \footnotesize  Including optimisation of hyperparameters.}\\
    \end{tabular}
\end{table}

\begin{figure}[h]
    \centering
    \includegraphics[width=1\linewidth]{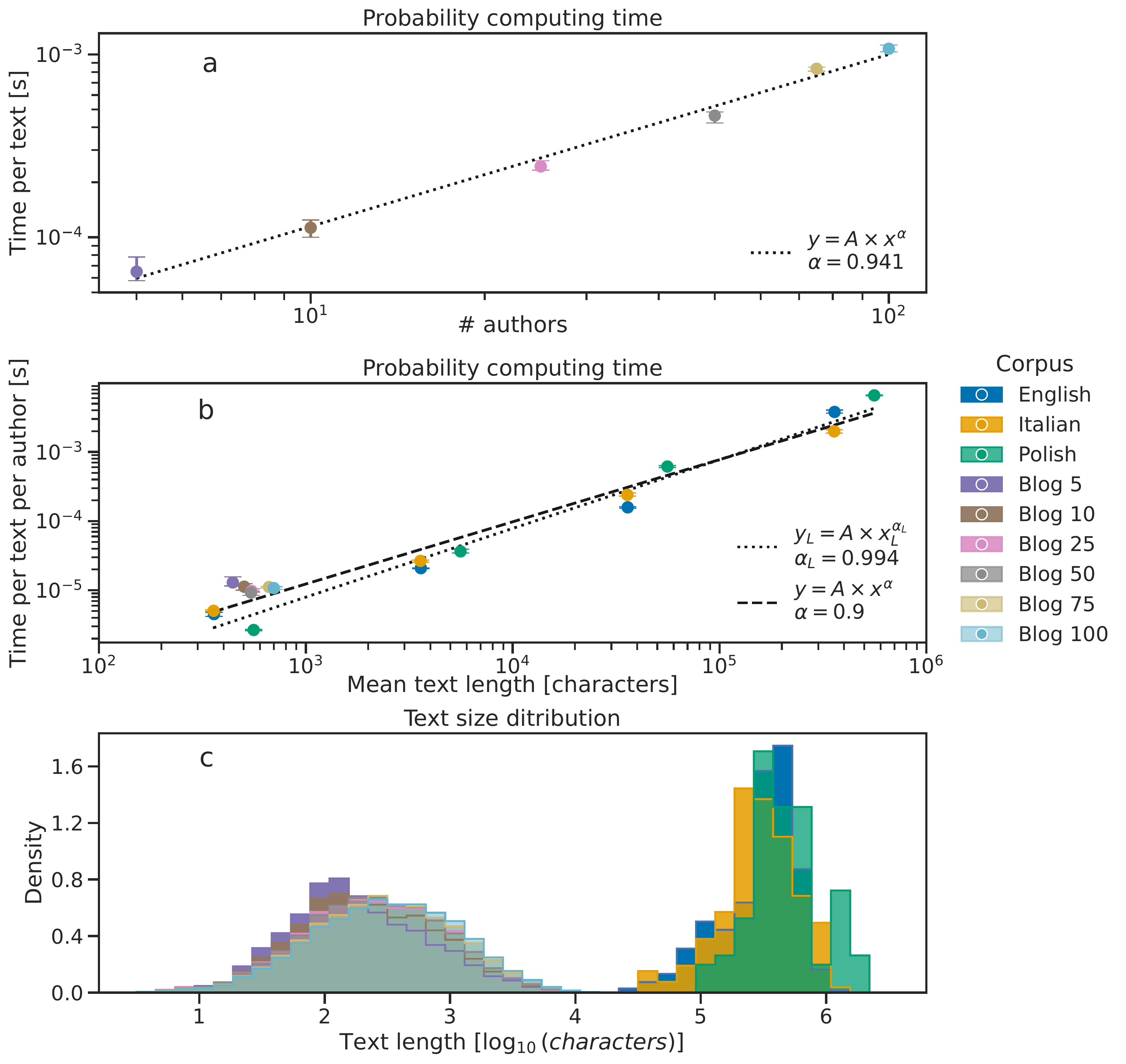}
    \caption{\textbf{Scaling laws of the execution time.} Panel (a): Average time to compute all the conditional probabilities of a text, i.e., conditional to each author, for the subset of the Blog corpus. Panel (a): Average time to compute the probability of a text conditional to one author for the subsets of the Blog corpus, the Literary corpora, and three artificial corpora obtained by dividing each document of the literary corpora into 10, 100, and 1000 parts. A fit is reported considering all the points and excluding those relative to the subsets of the Blog corpus ($y$ and $y_L$ respectively). Panel (b): Size distribution of the texts in subsets of the Blog corpus and in the literary corpora. The error bars in panels (a) and (b) mark the maximum variation of the execution time over 4 runs of the attribution on the same machine.}
    \label{fig:scaling_laws}
\end{figure}

Additionally, we show in Figure~\ref{fig:scaling_laws} two plots that focus on the running time of the core computational elements, not accounting for the time spent loading or saving files, loading libraries, and so on. While these elements can noticeably impact the total running times, their scaling depends more on the architecture and system and would blur the picture.

Panel A shows the average time taken to compute all the conditional probabilities of a text, once the parameters $\alpha$ and $\theta$ defining the authors are fixed, as a function of the number of authors.  In this case we attribute the entire document at once and report results  for the subsets of the Blog corpus, as they have a comparable distribution of post lengths (see panel C). This allows us to observe the approximately linear growth of the computing time as a function of the number of authors.
In panel B we give a glimpse of the dependence of the computing time on the text length. We report here
 the average time to compute the conditional probability of a text (i.e.,
the same observable as in panel A, additionally divided by the number of authors in the corpus)  as a function of the average text length in the corpus.
 We report results for the subsets of the Blog corpus and for the literary corpora. Additionally, to cover a broader range of average document sizes, we consider three artificial corpora obtained by dividing each document of the literary corpora
into 10, 100, and 1000 parts and consider each of those as texts. An approximately linear growth of the computing time as a function of the average document size can be observed.

\FloatBarrier

\bibliography{Supporting_natcomm_revised_2}